\documentclass{article}%
\usepackage{amsmath}
\usepackage{amsfonts}
\usepackage{amssymb}
\usepackage{graphicx}%
\begin{document}

\begin{center}
{\Large Proposal of a method for detecting dull images}
\end{center}

\medskip P. Brovetto$^{\dag}$(\footnote{Corresponding Author e-mail:
pbrovetto@gmail.com}, V. Maxia$^{\dag}$ and M. Salis$^{\dag\ddag}$

$^{\dag}$On leave from Istituto Fisica Superiore - University of Cagliari, Italy

$^{\ddag}$Dipartimento di Fisica - University of Cagliari, Italy

\textbf{Abstract - }Arguments are proposed which show how images hardly
perceptible as hided in a clouding backgound can be revealed utilizing special
patterns drawn by means of a simple mathematical procedure. Possible
applications might be found in medical radiology.


Digitally recorded images often require to be processed so as to remedy lacks
of readability due either to the limited capabilities of the recording devices
or to the intrinsic features of the images itself, for instance, a poor
contrast. Indeed, objects showing fading edges inevitably yield poor-contrast
images, but also objects caracterized by sharp edges, if buried in a heavy
background, cannot allow good-contrast images. Cases of this kind are common
in medical radiology. Organs in the human body, such as the blood vessels or
the digestive tracts, have indeed well-defined edges, but often X-ray opaque
matters must be there injected to allow their X-ray imagery. Other cases of
this kind happen in archaeology. Ancient masonries buried in the soil
sometimes can be detected if the dim contrast they originate in the aerial
pictures of archaeological sites is made out.

Focusing our attention on the just outlined problems of imagery of
hided-in-background sharp-edge objects, we consider the possibility of detect
these images utilizing special patterns drawn by means of a mathematical
procedure purposely devised (\footnote{) Data on conventional methods to
improve contrast in digital images can be found in Wikipedia Ref. 1.}.

Utilizing a scale of greys or false colors, any monochromatic image lying on
$x,y$ plane is defined by a positive function $\phi\left(  x,y\right)  $. When
variation of $\phi$ for variations of $x,y$ in the range of the significant
details of the image is very small with respect to actual values of $\phi$,
the image is dull, that is, barely perceptible. In practical cases, without
significant limiting of generality, function $\phi$ can be represented on
$x,y$ plane by means of contours, that is, lines where $\phi$ is constant. Let
$\Lambda$ be any contour and $O$ any point on this contour. We chose point $O$
as the origin of $x$, $y$ axes and also as the origin of two orthogonal
$\xi,\eta$ axes, the $\xi$ axis being tangent to line $\Lambda$.
Transformation between these axes is%

\[
\xi=\cos\alpha\cdot x+\sin\alpha\cdot y,
\]

\[
\eta=-\sin\alpha\cdot x+\cos\alpha\cdot y,
\]
$\alpha$ standing for the angle between $x$ and $\xi$ axes. By taking into
account that at the point $O$ we have $\partial\phi/\partial\xi=0$ and putting
$\partial\phi/\partial\eta=\delta^{\ast}$, we get%

\[
\frac{\partial\phi}{\partial x}=\frac{\partial\phi}{\partial\xi}%
.\frac{\partial\xi}{\partial x}+\frac{\partial\phi}{\partial\eta}%
.\frac{\partial\eta}{\partial x}=-\delta^{\ast}\cdot\sin\alpha,
\]

\[
\frac{\partial\phi}{\partial y}=\frac{\partial\phi}{\partial\xi}%
.\frac{\partial\xi}{\partial y}+\frac{\partial\phi}{\partial\eta}%
.\frac{\partial\eta}{\partial y}=\delta^{\ast}\cdot\cos\alpha,
\]
which leads to%

\begin{equation}
\delta\left(  x,y\right)  =\sqrt{\left(  \frac{\partial\phi}{\partial
x}\right)  ^{2}+\left(  \frac{\partial\phi}{\partial y}\right)  ^{2}},
\label{aa}%
\end{equation}
where $\delta=\left\vert \delta^{\ast}\right\vert $. Since function $\delta$
depends only on variation of $\phi$ on $x,y$ plane, it allows to reveal the
details of a dull image. To apply equation (\ref{aa}) derivatives
$\partial\phi/\partial x$ and $\partial\phi/\partial y$ must be evaluated
numerically. Assuming, for instance, that $\phi$ is known in equally-spaced
points with constant interval $h$ (pixels), derivative in the central point
among seven ones is%

\[
\phi_{4}^{\prime}=\left(  -\phi_{1}+9\phi_{2}-45\phi_{3}+45\phi_{5}-9\phi
_{6}+\phi_{7}\right)  /60h-h^{6}\phi^{\left(  7\right)  }/140,
\]
both on $x$ and $y$ axes. In the remainder term, $\phi^{\left(  7\right)  }$
means the seventh derivative of $\phi$ taken at some point interior to the $x$
or $y$ interval. Formulas for points out of centre are also available (Milne 1949).

In order to show how this procedure works, let us consider the simple example
of a disk of radius $R$ in which value of function $\phi$ barely exceeds that
in disk environment where only background exists. Accordingly, it is
convenient to split function $\phi$ in two contributions, that is,%

\[
\phi\left(  x,y\right)  =\phi_{D}\left(  x,y\right)  +\phi_{B}\left(
x,y\right)  .
\]
Assuming $D$ be an arbitrarily-small positive constant, the disk spoken of can
be represented by%

\begin{equation}
\phi_{D}\left(  x,y\right)  =\frac{D}{1+\exp\left[  \left(  \sqrt{x^{2}+y^{2}%
}-R\right)  /\Gamma_{D}\right]  }, \label{ab}%
\end{equation}
where $\Gamma_{D}$ is a length related to the steepness of the disk border
(\footnote{) For simplicity sake, in equation (\ref{ab}) we utilize a Fermi's
function. Alternatively, the more steep Erf function could be considered.}.
Indeed, by letting $r=\sqrt{x^{2}+y^{2}}$ and assuming $\Gamma_{D}\ll R$, in
the disk outside, that is, for $\left(  r-R\right)  /\Gamma_{D}\gg0$, we have
$\phi_{D}\simeq0$ and $\phi\simeq\phi_{B}$. On the disk contour, that is, for
$r=R$, we have $\phi=D/2+\phi_{B}$. Like, in the disk inside, that is, for
$\left(  r-R\right)  /\Gamma_{D}\ll0$, we have $\phi\simeq D+\phi_{B}$. So, if
$\phi_{B}$ is considered constant, it follows from (\ref{aa}) and (\ref{ab}) that%

\[
\delta\left(  r\right)  =\frac{D}{\Gamma_{D}}\frac{\exp\left[  \left(
r-R\right)  /\Gamma_{D}\right]  }{\left\{  1+\exp\left[  \left(  r-R\right)
/\Gamma_{D}\right]  \right\}  ^{2}}.
\]
It is easy to check that%

\begin{equation}
\delta\left(  R\right)  =\frac{D}{4\Gamma_{D}};\quad\quad\quad\left(
\frac{d\delta}{dr}\right)  _{r=R}=0;\quad\quad\quad\left(  \frac{d^{2}\delta
}{dr^{2}}\right)  _{r=R}=-\frac{D}{8\Gamma_{D}^{3}}, \label{bc}%
\end{equation}
which mean that function $\delta\left(  r\right)  $ shows a maximum on the
disk contour. We have moreover%

\[
\frac{D}{\Gamma_{D}}\frac{\exp\left[  \left(  r-R\right)  /\Gamma_{D}\right]
}{\left\{  1+\exp\left[  \left(  r-R\right)  /\Gamma_{D}\right]  \right\}
^{2}}\equiv\frac{D}{\Gamma_{D}}\frac{\exp\left[  \left(  R-r\right)
/\Gamma_{D}\right]  }{\left\{  1+\exp\left[  \left(  R-r\right)  /\Gamma
_{D}\right]  \right\}  ^{2}},
\]
which means that function $\delta\left(  r\right)  $ is symmetrical across its
maximum value $\delta\left(  R\right)  $. Utilizing equation%

\[
\frac{D}{\Gamma_{D}}\frac{\exp\left[  \sigma/\Gamma_{D}\right]  }{\left\{
1+\exp\left[  \sigma/\Gamma_{D}\right]  \right\}  ^{2}}=\frac{1}{2}\cdot
\frac{D}{4\Gamma_{D}},
\]
the HWHM $\sigma$ of function $\delta\left(  r\right)  $ is found to be%

\[
\sigma=1.763\cdot\Gamma_{D}.
\]
Consequently, it follows from (\ref{ab}) that function $\phi_{D}\left(
r\right)  $ increases from $0.146\cdot D$ to $0.853\cdot D$, for $r$
increasing from $R-\sigma$ to $R+\sigma.$

The found results show that the disk $\phi_{D}\left(  r\right)  $ corresponds
to a ring $\delta\left(  r\right)  $\ with the same radius $R$ of the disk,
the "thickness" $\sigma$ of this ring being related to the steepness
$\Gamma_{D}$ of the disk border. It is important to point out that a result of
this kind is fair for whatever figure on $x,y$ plane more complex than a disk,
provided that it can be defined by a contour line $\Lambda$ allowing in all
its points for a tangent $\xi$ axis. So, function $\delta$ is expected to
represent in general a contour pattern, with a "contour-thickness" $\sigma$
related to the sharpeness of the figure border. This result is based on the
assumption of $\phi_{B}$ constant. But even small variation of $\phi_{B}$ in
$x$, $y$ plane affect the actual value of $\delta$ so that readability of the
contour pattern might be worsed by a fragmentary backgound. A simple istance is%

\[
\phi_{B}\left(  x\right)  =B_{0}+B_{1}\frac{x}{\Gamma_{B}},
\]
$B_{0}$ and $B_{1}$ standing for constant quantities and $\Gamma_{B}$ for a
length large with respect to $\Gamma_{D}$ representing the variation of the
image background along $x$-axis. Owing to (\ref{aa}), we have%

\[
\delta_{B}=\frac{\partial\phi_{B}}{\partial x}=\frac{B_{1}}{\Gamma_{B}},
\]
so that the condition for the contour readability obviously is $\delta\left(
R\right)  \geq\delta_{B}$, that is, remembering (\ref{bc}),
\[
\frac{D}{4B_{1}}\geq\frac{\Gamma_{D}}{\Gamma_{B}}.
\]
If this condition is fulfilled, contour patterns $\delta$\ reveal the presence
of otherwise hardly perceptible dull images (\footnote{) To apply
transformation (\ref{aa}) when image is affected by a serious noise it could
be convenient utilize previously a \ proper noise-filtering \ procedure (see
Ref. 1).}. This allows conclude that transformation (\ref{aa}) could provide
data sufficient for characterizing the objects under consideration even when
their actual images are not available. In our opinion, the use of contour
patterns might have some interest especially in medical radiology.

\begin{center}
{\large References}
\end{center}

1) Wikipedia:How to improve image quality. http:en.wikipedia.org/wiki/.

2) W. E. Milne \textit{Numerical calculus} - Princeton University Press
(Princeton N. J. 1949) pag. 98.

\end{document}